\documentclass[journal]{IEEEtran}

\usepackage{soul}
\soulregister\cite7
\soulregister\ref7
\soulregister\pageref7

\usepackage{graphicx}
\ifCLASSINFOpdf
\else
\fi

\graphicspath{{figures/}}
\usepackage{tikz}

\usepackage{amsmath,amsfonts,stmaryrd,dsfont}
\usepackage{multirow}

\usepackage[linesnumbered,algoruled]{algorithm2e}

\usepackage{array}

\usepackage{url,cite}

\begin{document}

\title{Complex ISNMF: a Phase-Aware Model \\for Monaural Audio Source Separation}

\author{Paul~Magron,
	Tuomas~Virtanen,~\IEEEmembership{Senior Member,~IEEE}
    \thanks{P. Magron and T. Virtanen are with the Laboratory of Signal Processing, Tampere University of Technology, Finland (e-mail: firstname.lastname@tut.fi). The work of P. Magron was partly supported by the Academy of Finland, project no.\ 290190. }
}


\maketitle

\begin{abstract}
This paper introduces a phase-aware probabilistic model for audio source separation. Classical source models in the short-time Fourier transform domain use circularly-symmetric Gaussian or Poisson random variables. This is equivalent to assuming that the phase of each source is uniformly distributed, which is not suitable for exploiting the underlying structure of the phase. Drawing on preliminary works, we introduce here a Bayesian anisotropic Gaussian source model in which the phase is no longer uniform. Such a model permits us to favor a phase value that originates from a signal model through a Markov chain prior structure. The variance of the latent variables are structured with nonnegative matrix factorization (NMF). The resulting model is called complex Itakura-Saito NMF (ISNMF) since it generalizes the ISNMF model to the case of non-isotropic variables. It combines the advantages of ISNMF, which uses a distortion measure adapted to audio and yields a set of estimates which preserve the overall energy of the mixture, and of complex NMF, which enables one to account for some phase constraints. We derive a generalized expectation-maximization algorithm to estimate the model parameters. Experiments conducted on a musical source separation task in a semi-informed setting show that the proposed approach outperforms state-of-the-art phase-aware separation techniques.
\end{abstract}


\IEEEpeerreviewmaketitle

\section{Introduction}

\IEEEPARstart{T}{he} goal of audio source separation~\cite{Comon2010} is to extract underlying \textit{sources} that add up to form an observable audio \textit{mixture}. In this paper, we address the problem of \textit{monaural} source separation, which means that the observed audio signal has been recorded through a single microphone.

To tackle this issue, many techniques act on a time-frequency (TF) representation of the data, such as the short-time Fourier transform (STFT), since the structure of audio signals is more prominent in that domain. In particular, nonnegative matrix factorization (NMF)~\cite{Lee1999} techniques have shown successful for audio source separation~\cite{Virtanen2007,Fevotte2009}. NMF is a rank-reduction method used for obtaining part-based decompositions of nonnegative data. The NMF problem is expressed as follows: given a matrix $\textbf{V}$ of dimensions $F \times T$ with nonnegative entries, find a factorization $\textbf{V} \approx \textbf{WH}$ where $\textbf{W}$ and $\textbf{H}$ are nonnegative matrices of dimensions $F \times K$ and $K \times T$ respectively. To reduce the dimensionality of the data, the rank $K$ is generally chosen so that $K(F+T)\ll FT $. In audio applications $\textbf{V}$ is usually a magnitude or power spectrogram, and one can interpret $\textbf{W}$ as a dictionary of spectral templates and $\textbf{H}$ as a matrix of temporal activations.

Such a factorization is generally obtained by minimizing a cost function that penalizes the error between $\textbf{V}$ and $\textbf{WH}$. Popular choices are the Euclidean distance or Kullback-Leibler (KL) \cite{Lee1999} and Itakura-Saito (IS) divergences~\cite{Fevotte2009}. NMF may often be framed in a probabilistic framework, where the cost function appears as the negative log-likelihood of the data~\cite{Virtanen2008,Fevotte2009,Liutkus2015a,Simsekli2015}, and where the model structures the dispersion parameter of the underlying probability distribution rather than its observed realizations. For instance, in additive Gaussian mixtures~\cite{Fevotte2005} where the NMF models the variance of the sources, maximum likelihood estimation is equivalent to an NMF with IS divergence (ISNMF) of the power spectrogram~\cite{Fevotte2009}.

Once the NMF model has been estimated, the complex-valued STFTs are retrieved by means of a Wiener-like filter~\cite{Liutkus2015}. This soft-masking of the complex-valued mixture's STFT assigns the phase of the original mixture to each extracted source. However, even if this filter yields quite satisfactory sounding estimates in practice~\cite{Fevotte2009,Virtanen2007}, it has been pointed out~\cite{Magron2015} that when sources overlap in the TF domain, it is responsible for residual interference and artifacts in the separated signals. This is a consequence of assuming that the phase is uniformly distributed~\cite{Parry2007}, and therefore of not exploiting its underlying structure.

To alleviate this issue, the complex NMF (CNMF) model~\cite{Kameoka2009} has been proposed. It consists in directly decomposing the complex-valued mixture's STFT into a sum of rank-1 components whose magnitudes are structured by means of an NMF. This model allows for jointly estimating the magnitude and the phase of each source. It is estimated by minimizing the Euclidean distance between the model and the data, to which can be added some regularization terms, such as a sparsity penalty~\cite{Kameoka2009}. It was later improved by means of adding a \textit{consistency} constraint~\cite{LeRoux2009a}, that is, to account for the redundancy of the STFT which introduces some dependencies between adjacent TF bins~\cite{Griffin1984,LeRoux2008}.

Alternatively, improved recovery can be achieved by using phase constraints that originate from a signal model. For instance, the model of sums of sinusoids~\cite{McAuley1986} leads to explicit constraints between the phases of adjacent TF bins~\cite{Krawczyk2012,Magron2018}. Such an approach has been exploited in speech enhancement~\cite{Krawczyk2014,Mowlaee2015}, audio restoration~\cite{Magron2015a} and for a time-stretching application in the phase vocoder algorithm~\cite{Laroche1999}. It has also been incorporated into some phase-constrained CNMF models for audio source separation~\cite{Bronson2014,Rodriguez-Serrano2016,Magron2016}. Those developments have shown promising results in terms of interference rejection, though they suffer from two drawbacks. Firstly, the CNMF model is estimated by minimizing a Euclidean distance, which does not properly characterize the properties of audio (such as its large dynamic range), where alternative divergences (such as KL or IS) are preferred~\cite{Gray1980}. Secondly, the set of estimated sources does not preserve the overall energy of the mixture, which leads to artifacts in the separated signals.

Drawing on those observations, we proposed in a preliminary work~\cite{Magron2017} to model the sources with anisotropic Gaussian (AG) variables, i.e., where the phase is no longer uniform. In such a model, one can promote a phase value which is obtained by exploiting the sinusoidal model. Estimation in a minimum mean square error sense results in an anisotropic Wiener filter, which optimally combines the mixture phase and the underlying phase model. We further introduced in~\cite{Magron2018a} a general Bayesian framework in which both magnitudes and phases were modeled as random variables, and the sinusoidal model was promoted through a Markov chain prior structure on the phase location parameter. However, in those preliminary approaches, the variance parameters were left unconstrained and therefore either assumed known or estimated beforehand.

In this paper, we introduce a Bayesian AG model that overcomes the limitations of those approaches. We structure the variance parameters of the sources by means of an NMF model, so we can jointly estimate the magnitudes and the phases in a unified framework. This model, called \textit{complex ISNMF}, combines the benefits of both ISNMF and CNMF: 
\begin{enumerate}
\item It is phase-aware;
\item The set of estimators is \textit{conservative}, i.e., their sum is equal to the observed mixture;
\item The estimation is based on the minimization of an IS-like divergence, which is appropriate for audio~\cite{King2012b}.
\end{enumerate}
In order to infer the parameters of the model, we derive a generalized expectation-maximization (EM) algorithm. This model is applied to a musical source separation task in a semi-informed setting. It outperforms both the traditional phase-unaware ISNMF and the phase-constrained CNMF model~\cite{Magron2016}. This demonstrates the usefulness of such a phase-aware Bayesian AG model to perform the joint estimation of magnitudes and phases for audio source separation.

The rest of this paper is organized as follows. Section~\ref{sec:model} introduces the complex ISNMF model. Section~\ref{sec:estimation} details the inference procedure. Section~\ref{sec:exp} experimentally validates the potential of this method. Finally, Section \ref{sec:conclu} draws some concluding remarks.

\section{Complex ISNMF}
\label{sec:model}

Let $\textbf{X} \in \mathbb{C}^{F \times T}$ be the STFT of a single-channel audio signal, where $F$ and $T$ are the numbers of frequency channels and time frames. $\textbf{X}$ is the linear and instantaneous mixture of $J$ sources $\textbf{S}_j \in \mathbb{C}^{F \times T}$, such that for all TF bins $ft$,
\begin{equation}
x_{ft} = \sum_{j=1}^J s_{j,ft}.
\label{eq:mix}
\end{equation}
Since all TF bins are treated similarly, we remove the indices $ft$ when appropriate for more clarity.

\subsection{Modeling magnitude and phase}
\label{sec:model_rvm}

Let us consider a complex-valued random variable $s = r e^{\mathrm{i} \phi}$ whose magnitude and phase are assumed independent and denoted $r$ and $\phi$. Drawing on~\cite{Magron2018a}, we propose to model $r$ as a Rayleigh random variable $\mathcal{R}(v)$, which is the distribution of the modulus of a circularly-symmetric complex normal distribution with variance $v$. Besides, as in~\cite{Magron2017}, we consider that the phase should be distributed around some favored value $\mu$ and that the relative importance of this value should be adjusted by means of a concentration parameter $\kappa \in [0,+\infty[$: the higher $\kappa$, the more favored $\mu$.

\begin{figure}[t]
	\centering
	\includegraphics[scale=0.45]{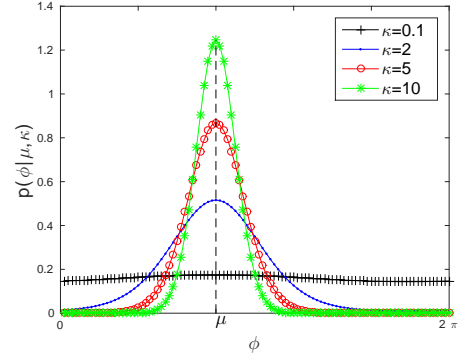}
	\caption{Density of the VM distribution.}
	\label{fig:vonmises_density}
\end{figure}

Several non-uniform periodic distributions exist (such as the wrapped Gaussian~\cite{Agiomyrgiannakis2009} or wrapped Cauchy distributions) but the von Mises (VM)~\cite{Mardia1975} distribution comes as a natural candidate~\cite{Gerkmann2014,Gerkmann2014a}, since its density is easily expressed by:
\begin{equation}
p(\phi|\mu,\kappa) = \displaystyle \frac{e^{  \kappa \cos (\phi - \mu)}}{{2 \pi I_0(\kappa)}},
\end{equation}
where $I_n$ is the modified Bessel function of the first kind of order $n$~\cite{Watson1995}, $\mu \in [0;2\pi[$ is a location parameter and $\kappa \in [0;+\infty[$ is a concentration parameter. In particular, if $\kappa = 0$, the VM distribution becomes uniform. Contrarily, if $\kappa \to +\infty$, it becomes equivalent to a Dirac delta function centered at $\mu$. It is illustrated in Fig.~\ref{fig:vonmises_density}.
         
This methodology results in a model called Rayleigh + von Mises (RVM), in which one can promote some favored phase values (see Section~\ref{sec:model_phase}). Such an approach has been originally used in~\cite{Gerkmann2014,Gerkmann2014a} for a speech enhancement application in a speech plus noise model. However, in the present case, since we consider any number of sources $J$, the RVM model is no longer tractable because the density of the mixture does not admit a closed-form expression. Therefore it is not suitable for source separation, where we aim to estimate the model parameters.

Nonetheless, we can compute the moments of $s=r e^{\mathrm{i} \phi}$ which will be used later in this work. If $\phi \sim \mathcal{VM}(\mu,\kappa)$, the $n$-th circular moment is, $\forall n \in \mathbb{Z}$ (\textit{cf.}~\cite{Mardia1975}):
\begin{equation}
\mathbb{E}(e^{\mathrm{i} n \phi}) = \displaystyle \frac{I_{|n|}(\kappa)}{I_{0}(\kappa)} e^{\mathrm{i} n \mu}.
\end{equation}
Besides, if magnitude $r \sim \mathcal{R}(v)$, we have:
\begin{equation}
\mathbb{E}(r) = \sqrt{\frac{\pi}{4} v} \text{ and } \mathbb{E}(r^2) = v.
\end{equation}
This lead to the expression of the mean of $s$:
\begin{equation}
m =\mathbb{E}(re^{\mathrm{i} \phi}) = \mathbb{E}(r) \mathbb{E}(e^{\mathrm{i} \phi}) = \lambda \sqrt{v} e^{\mathrm{i} \mu},
\label{eq:rvm_mean}
\end{equation}
and its variance $\gamma = \mathbb{E}(|s-m|^2)$:
 \begin{equation} 
\gamma = \mathbb{E}(|re^{\mathrm{i} \phi}|^2) - |m|^2 =  (1- \lambda^2 ) v,
\label{eq:rvm_var}
\end{equation}
and the relation term $c = \mathbb{E}((s-m)^2)$, which measures the joint variability of a variable and its complex conjugate:
\begin{equation} 
c  = \mathbb{E}(r^2) \mathbb{E}(e^{\mathrm{i} 2 \phi}) - m^2 = \rho  v  e^{\mathrm{i} 2\mu},  \label{eq:rvm_rel}
\end{equation}
where
\begin{equation}
\lambda = \displaystyle \frac{ \sqrt{\pi}}{2} \frac{I_{1}(\kappa)}{I_{0}(\kappa)} \text{ and } \rho = \displaystyle \frac{I_{2}(\kappa)}{I_{0}(\kappa)} - \lambda^2.
\label{eq:ag_params}
\end{equation}
This relation term $c$ is not commonly introduced in statistical models of audio signals in the TF domain because it is usually assumed to be null~\cite{Picinbono1996}. Indeed, most models~\cite{Fevotte2009,Liutkus2015,Liutkus2018} assume the second-order circularity (or \textit{isotropy}) of the variables, that is, with the same distribution in the complex plane regardless of the orientation. Since this is equivalent to assuming that the phase is uniformly distributed, we propose instead to explicitly consider this relation term as non-zero in general: it enables us to promote the non-circularity of the variable, and therefore the non-uniformity of the phase.

\subsection{Anisotropic Gaussian sources}
\label{sec:model_AG}

To alleviate the non-tractability issue of the RVM model, we propose to approximate it by a Gaussian model\footnote{This strategy is reminiscent of~\cite{Beckmann1962}, where the mixture model was a sum of random variables with phase priors.} in which the moments of the variables are the same ones as in the original RVM model. This approach enables us to keep the phase dependencies in a model which is fully tractable.

Therefore, we assume that each source $s_j$ follows a complex normal distribution: $s_j \sim \mathcal{N}(m_j,\Gamma_j)$, where $m_j= \mathbb{E}(s_j) \in \mathbb{C}$ is the mean of $s_j$ and $\Gamma_j$ is its covariance matrix:
\begin{equation}
\Gamma_j = 
\begin{pmatrix}

\gamma_j & c_j \\
\bar{c}_j & \gamma_j

\end{pmatrix},
\end{equation}
where $\gamma_j= \mathbb{E}(|s_j-m_j|^2) \in \mathbb{R}_+$ and $c_j = \mathbb{E}((s_j-m_j)^2) \in \mathbb{C} $ are the variance and relation term of $s_j$, and $\bar{z}$ denotes the complex conjugate of $z$. The density of such a distribution is:
\begin{equation}
p(x|m,\Gamma) = \displaystyle \frac{1}{\pi \sqrt{| \Gamma|}}  e^{- \frac{1}{2} ( \underline{x}-\underline{m} )^\mathsf{H}   \Gamma^{-1} ( \underline{x}-\underline{m} ) },
\end{equation}
where $\underline{x} = \begin{pmatrix} x & \bar{x} \end{pmatrix}^\mathsf{T}$, and where $^\mathsf{T}$ and $^\mathsf{H}$ denote the transpose and conjugate transpose. 

\begin{figure}[t]
	\includegraphics[scale=0.75]{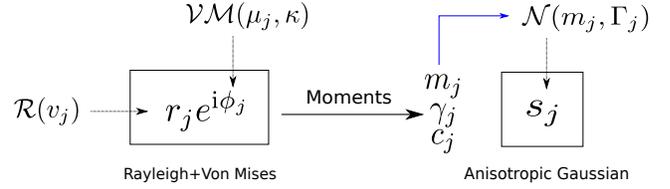}
	\caption{Design of the AG model. We first model the magnitudes and phases as Rayleigh and von Mises random variables. The moments in this model are then used to define the equivalent AG model.}
	\label{fig:rvm_ag}
\end{figure}

Many previous studies model the sources as circularly-symmetric (or \textit{isotropic}) variables~\cite{Fevotte2009,LeRoux2013} (i.e., such that $m_j=c_j=0$), which is equivalent to assuming that the phase of each source is uniformly distributed. The keystone of our approach is that, in order to promote a favored phase value, the moments are the same ones as in the original RVM model. Therefore, we use the expressions given by~\eqref{eq:rvm_mean},~\eqref{eq:rvm_var} and~\eqref{eq:rvm_rel} to estimate the moments which are then used to design the Gaussian model, as illustrated in Fig.~\ref{fig:rvm_ag}. The main characteristic of this model is that the relation terms $c_j$ are non-zero in general, which conveys the property of \textit{anisotropy} of the corresponding Gaussian distribution: this is why we refer to it as the anisotropic Gaussian (AG) model.

The additive property of the Gaussian distribution family then implies that $x \sim \mathcal{N}(m_x,\Gamma_x)$ with:
\begin{equation}
m_x=\sum_j m_j \text{, } \gamma_x=\sum_j \gamma_j \text{, } c_x=\sum_j c_j \text{, } \Gamma_x=\sum_j \Gamma_j.
\label{eq:moments_X}
\end{equation}

\begin{figure}[t]
	\centering
	\includegraphics[scale=0.55]{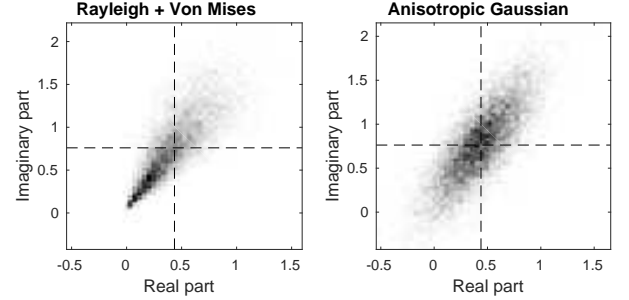}
	\caption{2-D histograms of $10000$ samples generated from the RVM model (left) and AG model (right), with $v=1$, $\mu=\pi/3$ and $\kappa=50$. The intersection between the dashed lines represents the mean of the samples.}
	\label{fig:hist_vm_ag}
\end{figure}

\noindent \textbf{Remark}: If $\kappa = 0$, then $\lambda = \rho = 0$ and consequently $m = c = 0$ and $\gamma = v$: the RVM and AG models are then equivalent since they both become isotropic Gaussian. Contrarily, for important values of $\kappa$, the models still remain quite alike, as illustrated in Fig.~\ref{fig:hist_vm_ag} for $\kappa = 50$.

\subsection{Phase model}
\label{sec:model_phase}

The non-uniformity of the phase is taken into account in the AG model through the location parameter $\mu$. However, in order to obtain good quality phase estimates, this model can benefit from incorporating some prior knowledge about the phase, for instance by accounting for its structure in time or frequency.
We propose to exploit some information about the phase by exploiting the sinusoidal model, which is widely used for representing audio signals~\cite{Krawczyk2014,Bronson2014}. Each source in the time domain is modeled as a sum of sinusoids. Let us assume that there is at most one sinusoid (whose normalized frequency is denoted $\nu_{j,ft}$) per frequency channel. It can be shown~\cite{Magron2015a} that the phase $\mu_j$ follows the unwrapping equation:
\begin{equation}
\mu_{j,ft} \approx \mu_{j,ft-1} + 2 \pi l \nu_{j,ft},
\label{eq:phase_unwrapping}
\end{equation}
where $l$ is the hop size of the STFT. As in~\cite{Magron2018a}, we propose to enforce this property by means of a Markov chain prior structure. We have, for each source:
\begin{equation}
p(\mu_j) =  \prod_{f = 0}^{F-1} p(\mu_{j,f0})  \prod_{t = 1}^{T-1}   p(\mu_{j,ft}| \mu_{j,ft-1} ).
\end{equation}
We then propose the following choice, for $t > 0$:
\begin{equation}
\mu_{j,ft} | \mu_{j,ft-1} \sim  \mathcal{VM}(\mu_{j,ft-1}+2 \pi l \nu_{j,ft},\tau),
\label{eq:prior_mu}
\end{equation}
and the initial distribution in each frequency channel $p(\mu_{j,f0})$ is Jeffrey's non-informative prior. In this way, we enforce the phase location parameter to approximately follow the sinusoidal model~\eqref{eq:phase_unwrapping}. The parameter $\tau \in \mathbb{R}_+$ adjusts the relative importance of this prior.
Once again, we choose a VM distribution for modeling the phase location parameter, since it is a natural candidate for accounting for the periodicity of this variable. However, unlike previously, we do not need here to approximate this distribution: since the prior~\eqref{eq:prior_mu} applies independently to each source, it is straightforward to explicitly obtain the log-prior:
\begin{equation}
\log ( p(\boldsymbol{\mu}) ) \overset{c}{=}  \tau \sum_{j,f,t} \Re \left( e^{\mathrm{i} \mu_{j,ft}} e^{- \mathrm{i} \mu_{j,ft-1} -2 \mathrm{i} \pi l \nu_{j,ft}}   \right),
\label{eq:phaseprior}
\end{equation}
where $\overset{c}{=}$ denotes equality up to an additive constant and $\Re$ is the real part. The model therefore depends on two concentration parameters that have a different role: $\kappa$ quantifies the non-uniformity of the phase in the AG model (i.e., how concentrated about a location parameter the phase is), while $\tau$ quantifies how close to the sinusoidal model this location parameter is.

\subsection{Complex ISNMF}
\label{sec:model_nmf}

\begin{figure}[t]
	\centering
	\includegraphics[scale=0.45]{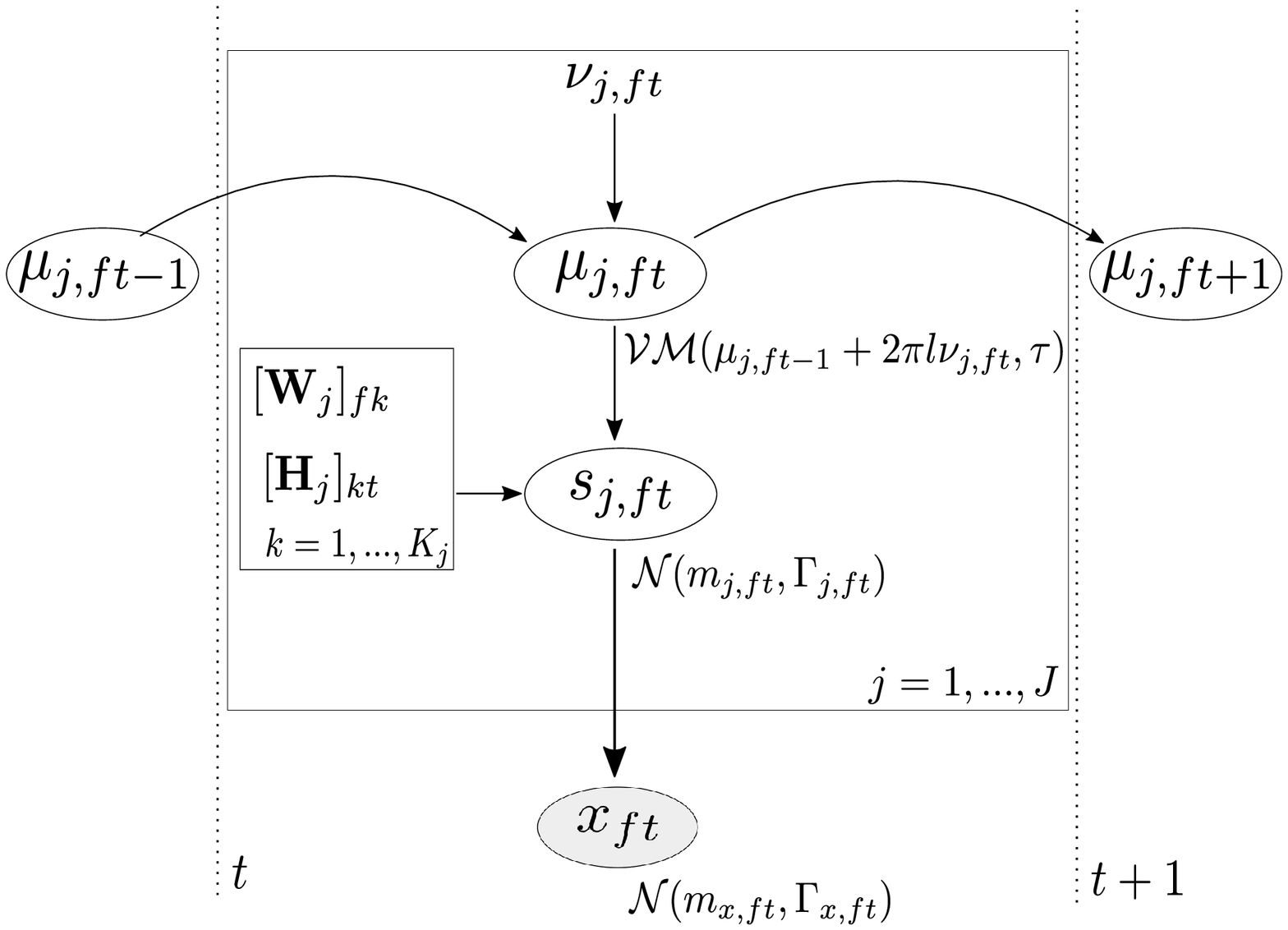}
	\caption{Bayesian network corresponding to the complex ISNMF model. Latent (resp. observed) variables are represented with empty (resp. shaded) ellipses. The sub-graph contained in each rectangle is repeated according to the index ($k$ or $j$) indicated in the bottom-right corner of the rectangle. The vertical dashed lines mark the limits between successive time frames.}
	\label{fig:cisnmf_network}
\end{figure}

For practical separation applications, it is necessary to constrain the variance parameters of the sources $\mathbf{V}_j$. We propose to structure it by means of an NMF model:
\begin{equation}
\mathbf{V}_j = \mathbf{W}_j \mathbf{H}_j,
\end{equation}
where $\mathbf{W}_j$ and $\mathbf{H}_j$ are nonnegative-valued matrices of dimensions $F \times K_j$ and $K_j \times T$ respectively. Therefore, the moments in the AG model become:
\begin{align}
m_{j,ft} &= \lambda  \sqrt{[\mathbf{W}_j \mathbf{H}_j]_{ft}} \text{ } e^{\mathrm{i} \mu_{j,ft}}, \nonumber\\
\gamma_{j,ft} &=  (1- \lambda^2 ) [\mathbf{W}_j \mathbf{H}_j]_{ft}, \label{eq:ag_moments} \\
c_{j,ft} &= \rho  [\mathbf{W}_j \mathbf{H}_j]_{ft}  \text{ } e^{\mathrm{i} 2\mu_{j,ft}}, \nonumber
\end{align}
where $[\mathbf{W}_j \mathbf{H}_j]_{ft}$ denotes the $(f,t)$-th entry of the matrix $\mathbf{W}_j \mathbf{H}_j$.
In particular, if $\kappa=0$, then $m_j=c_j=0$ and $\gamma_j =\mathbf{W}_j \mathbf{H}_j $: the model becomes equivalent to ISNMF. Thus, since the proposed model generalizes ISNMF while allowing us to account for some phase constraint, we call it \textit{complex ISNMF}. The whole model is represented as a Bayesian network in Fig.~\ref{fig:cisnmf_network}

\subsection{Relation to other models}
\label{sec:model_relation}

The AG model along with the NMF variance structure results in a phase-aware extension of ISNMF, as pointed out in Section~\ref{sec:model_nmf}. However, other models can be seen as particular cases of this general framework. Indeed, in Section~\ref{sec:model_AG} we approximated the RVM model with an AG model by equating their moments. As illustrated in Fig.~\ref{fig:rvm_ag}, we chose to equate all the moments (mean, variance and relation term), but other approaches are possible.

Firstly, it is possible to set the mean and relation term to $0$, in which case the sources follow a circularly-symmetric Gaussian distribution: $s_j \sim \mathcal{N}(0,\gamma_j I)$, where $I$ is the identity matrix. Along with an NMF variance, this results in the ISNMF model~\cite{Fevotte2009}. This is therefore another way of seeing the proposed AG model as an extension of ISNMF.

Alternatively, one can only preserve the mean information from the RVM model, and set the covariance matrix to be diagonal with a constant variance $\sigma$: $s_j \sim \mathcal{N}(m_j,\sigma I)$. This is the underlying statistical model from CNMF~\cite{Kameoka2009}. Therefore, this AG framework bridges the gap between ISNMF and CNMF since it generalizes both of them in a unified model.

Finally, other approximations are possible. For instance, one can only preserve the second-order statistics from the RVM model and set the mean value at 0 ($s_j \sim \mathcal{N}(0,\Gamma_j)$). Instead, one can set the relation terms at 0 and keep the phase dependencies only through the mean ($s_j \sim \mathcal{N}(m_j,\gamma_j I)$). This leads to alternative versions of Complex ISNMF that simplify the estimation of the NMF parameters (\textit{cf}. Section~\ref{sec:estim_nmf}) or the phase parameters (\textit{cf}. Section~\ref{sec:estim_phase}). Those will be discussed in the corresponding sections. However, in order to keep the scope of this paper broad enough, we will infer the model in the general case described in Section~\ref{sec:model_nmf}.

\section{Inference}
\label{sec:estimation}

The model parameters $\Theta = \{ \{\mathbf{W}_j \}_j, \{\mathbf{H}_j \}_j,  \{\mu_j\}_j  \}$ are estimated in a maximum a posteriori sense, which consists in maximizing the log-posterior distribution:
\begin{equation}
\mathcal{C}_{\text{MAP}}(\Theta) = \log p(\textbf{X}|\Theta) + \log p(\Theta),
\label{eq:CMAP}
\end{equation}
where $p(\textbf{X}|\Theta)$ is the likelihood of the data and $p(\Theta)$ the priors on the parameters. In this work, we only exploit the Markov prior information about the phase, therefore $\log p(\Theta)$ is given by~\eqref{eq:phaseprior}. However, this framework is very general and it could be possible to further enforce some desirable property such as harmonicity~\cite{Bertin2010} through priors on the columns of $\mathbf{W}_j$ or temporal continuity~\cite{Virtanen2007} through priors on the rows of $\mathbf{H}_j$.

\subsection{EM framework}

Since the direct maximization of the criterion~\eqref{eq:CMAP} is more involved than in classical isotropic models~\cite{Fevotte2009}, we propose to adopt an EM~\cite{Dempster1977} strategy which consists in maximizing a lower bound of the log-posterior distribution, given by:
\begin{equation}
\mathcal{Q}^{\text{MAP}}(\Theta,\Theta^{(i-1)}) = \mathcal{Q}^{\text{ML}}(\Theta,\Theta^{(i-1)}) + \log p(\Theta),
\label{eq:QMAP}
\end{equation}
where $i$ is a step index, $\Theta^{(i-1)}$ contains the current set of estimated parameters (i.e., the parameters estimated at the previous step $i-1$) and $\mathcal{Q}^{\text{ML}}$ is the conditional expectation of the complete-data log-likelihood:
\begin{equation}
\mathcal{Q}^{\text{ML}}(\Theta,\Theta^{(i-1)}) = \int p(\textbf{Z}|\textbf{X};\Theta^{(i-1)}) \log p(\textbf{X},\textbf{Z};\Theta)  d\textbf{Z},
\label{eq:QML}
\end{equation}
where $\textbf{Z}$ denotes a set of latent (hidden) variables. Due to the mixing constraint~\eqref{eq:mix}, we use, as in~\cite{LeRoux2013,Magron2017c}, a reduced set of $J'=J-1$ free variables $\textbf{Z} = \textbf{S} = \{ \textbf{s}_{ft} \}_{ft}$, where we note $\textbf{s}_{ft} = [s_{1,ft},...,s_{J',ft}]^\mathsf{T}$. Therefore, $s_{J,ft} = x_{ft}-\sum_{j=1}^{J'} s_{j,ft}$.

The EM algorithm consists in alternatively computing the functional $\mathcal{Q}^{\text{MAP}}$ given the current set of parameters $\Theta^{(i-1)}$ (E-step) and maximizing it with respect to $\Theta$ (M-step). This is proven~\cite{Dempster1977} to increase the value of the criterion~\eqref{eq:CMAP}. However, when the maximization of $\mathcal{Q}^{\text{MAP}}$ is too involved, it may be preferable to solely increase its value at the M-step. This has also been proved~\cite{Dempster1977} to lead to a local maximum of~\eqref{eq:CMAP}, and the corresponding procedure is called \textit{generalized} EM. This is the approach we are adopting hereafter.

\subsection{E-step}
\label{sec:estep}

Since all $\{s_{j,ft} \}_{j=1}^{J'}$ are independent Gaussian variables, $\textbf{s}_{ft}$ is a Gaussian vector.
It can be shown~\cite{Picinbono1996} that $\textbf{S}|\textbf{X}$ follows a multivariate complex normal distribution $\mathcal{N}(\textbf{m}'_{ft},\boldsymbol{\Xi}_{ft})$. The posterior means of the sources are given by anisotropic Wiener filtering~\cite{Magron2017}:
\begin{equation}
\underline{m}'_{j,ft}  = \underline{m}_{j,ft}^{(i-1)}   + \Gamma_{j,ft}^{(i-1)} \left( \Gamma_{x,ft}^{(i-1)} \right)^{-1} (\underline{x}_{ft} - \underline{m}_{x,ft}^{(i-1)} ).
\label{eq:posterior_mean}
\end{equation}
Note that, given the mixing constraint~\eqref{eq:mix}, this expression is also valid for the last source for which $j=J$. The posterior covariance matrix $\boldsymbol{\Xi}_{ft}$ is given by~\cite{Magron2017c}:
\begin{multline}
\boldsymbol{\Xi}_{ft} =
\begin{pmatrix}
\Gamma_{1,ft}^{(i-1)} & 0    & 0 \\
0         & \ddots    & 0   \\
0        & 0     & \Gamma_{J',ft}^{(i-1)}
\end{pmatrix}  \\
-
\begin{pmatrix}
\Gamma_{1,ft}^{(i-1)} \\ 
\vdots    \\
\Gamma_{J',ft}^{(i-1)}
\end{pmatrix}
\left( \Gamma_{x,ft}^{(i-1)} \right)^{-1} 
\begin{pmatrix}
\Gamma_{1,ft}^{(i-1)} \\ 
\vdots    \\
\Gamma_{J',ft}^{(i-1)}
\end{pmatrix}^\mathsf{T}.
\end{multline}
In particular, the diagonal blocks in the posterior covariance matrix provide the posterior covariance for each source:
\begin{equation}
\Gamma_{j,ft}' =  \Gamma_{j,ft}^{(i-1)}  - \Gamma_{j,ft}^{(i-1)} \left( \Gamma_{x,ft}^{(i-1)} \right)^{-1}  \Gamma_{j,ft}^{(i-1)}.
\label{eq:posterior_var}
\end{equation}
Thanks to~\eqref{eq:posterior_mean} and~\eqref{eq:posterior_var}, we can compute the posterior mean, variance and relation term of the sources, respectively, denoted by $m_j'$, $\gamma_j'$ and $c_j'$.
The computation of~\eqref{eq:QML} is detailed in the appendix and results in:
\begin{align}
\mathcal{Q}^{\text{ML}}(\Theta,&\Theta^{(i-1)}) \overset{c}{=} -\sum_{f,t}\sum_{j=1}^{J} \log(\sqrt{|\Gamma_{j,ft}|}) \nonumber \\
&+ \frac{1}{|\Gamma_{j,ft}|} \left( \gamma_{j,ft} (|m_{j,ft}'-m_{j,ft}|^2 + \gamma_{j,ft}')   \right) \label{eq:QMLlong} \\
&- \frac{1}{|\Gamma_{j,ft}|} \left( \Re( \bar{c}_{j,ft} ((m_{j,ft}'-m_{j,ft})^2 + c_{j,ft}') )  \right), \nonumber
\end{align}
where $|\Gamma_{j,ft}|=\gamma_{j,f,t}^2-|c_{j,ft}|^2$ is the determinant of $\Gamma_{j,ft}$.

\subsection{M-step: NMF parameters}
\label{sec:estim_nmf}

\subsubsection{NMF functional}
\label{sec:estim_nmf_func}

Let us first rewrite $\mathcal{Q}^{\text{ML}}$ by removing the terms that do not depend on the NMF parameters. Using~\eqref{eq:QMLlong} and~\eqref{eq:ag_moments}, we have:
\begin{align}
\mathcal{Q}^{\text{ML}}(\Theta|\Theta^{(i-1)}) \overset{c}{=} -\sum_{j=1}^J \sum_{f,t} & \log( [\mathbf{W}_j \mathbf{H}_j]_{ft}) + \frac{p_{j,ft}}{[\mathbf{W}_j \mathbf{H}_j]_{ft}} \nonumber \\
&- \frac{q_{j,ft}}{\sqrt{[\mathbf{W}_j \mathbf{H}_j]_{ft}}}, \label{eq:QMLnmf}
\end{align}
with:
\begin{equation}
p = \frac{ (1-\lambda^2) \left( \gamma' + |m'|^2 \right) - \rho \displaystyle \Re \left( e^{-2 \mathrm{i} \mu} (c' + m'^2 )   \right)}{(1-\lambda^2)^2-\rho^2},
\label{eq:p_em}
\end{equation}
and:
\begin{equation}
q = \frac{ 2 \lambda }{1-\lambda^2+\rho} \Re \left( e^{- \mathrm{i} \mu}  m'  \right),
\label{eq:q_em}
\end{equation}
where we removed the indices $j,ft$ for brevity. This highlights two novel quantities $p$ and $q$ on which $\mathcal{Q}^{\text{ML}}$ depends. First, from the derivation conducted in the appendix we remark that:
\begin{equation}
\frac{p_{j,ft}}{[\mathbf{W}_j \mathbf{H}_j]_{ft}} = \mathbb{E}_{\textbf{S}|\textbf{X};\Theta^{(i-1)}} \left( \underline{s}_{j,ft} ^\mathsf{H}   \Gamma_{j,ft}^{-1} \underline{s}_{j,ft}  \right).
\label{eq:p_esp}
\end{equation}
In particular, when $\kappa=0$, $p_{j,ft}= \gamma_{j,ft}' + |m_{j,ft}'|^2 $, which is the posterior power of $s_{j,ft}$. Therefore, in the general case, we call the quantity $p$ in~\eqref{eq:p_esp} the \textit{phase-corrected posterior power} of the sources. Note that since $\Gamma$ is positive-definite, $p$ is necessarily nonnegative. This quantity is interesting because it accounts for the phase while being nonnegative: therefore, estimating the NMF model from this quantity leads to a phase-aware decomposition of the data.

On the other hand, the physical meaning of the quantity $q$ is not fully clear. In particular, it has the same sign as  $\Re \left( e^{- \mathrm{i} \mu}  m' \right)$, that is, the same sign as $\cos (\mu - \angle m')$. Accounting for the mixture's phase when computing the posterior mean~\eqref{eq:posterior_mean} leads to a deviation of $\angle m'$ from the location parameter $\mu$. However, our intuition is that the posterior mean angle will stay relatively close to the location parameter $\mu$. If this angle difference remains relatively small (that is, $|\mu-\angle m'|<\pi/2$), then its cosine (and consequently $q$) is nonnegative. Then, $q$ has the dimension of a magnitude, and can therefore be seen as a \textit{phase-corrected posterior magnitude}. Even though we were not able to formally demonstrate that this intuition holds, we observed experimentally that $q$ was always nonnegative. Therefore, we will assume in what follows that $q$ is nonnegative, and we leave to future work a more in-depth analysis of those quantities.

\subsubsection{Majorize-minimization approach}

Since $\mathcal{Q}^{\text{MAP}}$ is equal to $\mathcal{Q}^{\text{ML}}$ up to the log-prior on the phase, which does not depend on the NMF parameters, the problem then becomes that of minimizing the following function, for all sources $j$:
\begin{equation}
\mathcal{H}(\Theta) = \sum_{f,t} \log( \sum_{k} w_{fk} h_{kt} ) + \frac{p_{ft}}{\sum_{k} w_{fk} h_{kt} } - \frac{q_{ft}}{\sqrt{\sum_{k} w_{fk} h_{kt}}}.
\label{eq:hnmf}
\end{equation}
To do so, we propose to adopt a majorize-minimization approach~\cite{Hunter2004}. The core idea of this strategy is to find an auxiliary function $\mathcal{G}$ which majorizes $\mathcal{H}$:
\begin{equation}
\forall (\Theta,\widetilde{\Theta}) \text{, } \mathcal{H}(\Theta) \leq \mathcal{G}(\Theta,\widetilde{\Theta}) \text{, and } \mathcal{H}(\widetilde{\Theta}) = \mathcal{G}(\widetilde{\Theta},\widetilde{\Theta}).
\label{eq:condMM}
\end{equation}
Given some current parameter $\widetilde{\Theta}$, minimizing $\mathcal{G}(\Theta,\widetilde{\Theta})$ with respect to $\Theta$ provides an update on $\Theta$. This approach guarantees that the cost function $\mathcal{H}$ is non-increasing over iterations.

Let us derive the update on $\textbf{W}_j$. We introduce auxiliary parameters $\widetilde{w}_{fk}$ and we denote $\widetilde{v}_{ft} = \sum_k \widetilde{w}_{fk} h_{kt}$. In a similar fashion as in~\cite{Fevotte2011,Fevotte2011a,Lefevre2011}, we decompose the function $\mathcal{H}$ into its convex and concave parts.

Since $p$ is nonnegative, the term in~\eqref{eq:hnmf} involving $p$ is convex. Therefore it is majorized by using the Jensen inequality:
\begin{equation}
\frac{p_{ft}}{\sum_{k} w_{fk} h_{kt} }  \leq  \sum_{k} \frac{\widetilde{w}_{fk}^2}{w_{fk}} \frac{p_{ft}h_{kt}}{\widetilde{v}_{ft}^2}.
\label{eq:maj1}
\end{equation}
Besides, since we assumed that $q$ is negative, the term in~\eqref{eq:hnmf} involving $q$ is concave, so it is majorized by its tangent:
\begin{equation}
-\frac{q_{ft}}{\sqrt{\sum_{k} w_{fk} h_{kt}}}  \leq  \sum_{k} \frac{w_{fk} h_{kt} q_{ft}}{\widetilde{v}_{ft}^{3/2}}.
\label{eq:maj2}
\end{equation}
Finally, the first term in~\eqref{eq:hnmf} is majorized as in~\cite{Fevotte2011a}:
\begin{equation}
\log ( \sum_{k} w_{fk} h_{kt} )   \leq  \sum_{k} \frac{w_{fk} h_{kt}}{\widetilde{v}_{ft}}.
\label{eq:maj3}
\end{equation}
Combining~\eqref{eq:maj1},~\eqref{eq:maj2} and~\eqref{eq:maj3} results into the following auxiliary function for $\mathcal{H}$:
\begin{equation}
\mathcal{G}(\Theta,\widetilde{\Theta}) =  \sum_{f,k} \frac{\widetilde{w}_{fk}^2}{w_{fk}} \sum_{t} \frac{p_{ft}h_{kt}}{\widetilde{v}_{ft}^2} + w_{fk} \sum_t h_{kt} ( \frac{1}{\widetilde{v}_{ft}} + \frac{q_{ft}}{\widetilde{v}_{ft}^{3/2}}).
\end{equation}

\subsubsection{Update rules}

Setting the derivative of $\mathcal{G}$ with respect to $w_{fk}$ at zero and solving leads to the following update:
\begin{equation}
w_{fk} = \widetilde{w}_{fk} \sqrt{\frac{\displaystyle \sum_t  \frac{p_{ft}h_{kt}}{\widetilde{v}_{ft}^2}}{\displaystyle \sum_t h_{kt} \left( \frac{1}{\widetilde{v}_{ft}} + \frac{q_{ft}}{\widetilde{v}_{ft}^{3/2}} \right) }}.
\end{equation}
We can rewrite this update rule onto matrix form as:
\begin{equation}
\mathbf{W}_j \leftarrow \mathbf{W}_j \odot \left( \frac{ (\mathbf{P}_j \odot \mathbf{V}_j ^{\odot -2}) \mathbf{H}_j^\mathsf{T}   }{(\mathbf{V}_j^{\odot -1} + \mathbf{Q}_j \odot \mathbf{V}_j^{\odot -3/2} ) \mathbf{H}_j^\mathsf{T} }     \right)^{\odot 1/2},
\label{eq:Wj_em}
\end{equation}
where $\odot$, $^{\odot}$ and the fraction bar denote element-wise matrix multiplication, power and division respectively, and where $\mathbf{P}_j$ and $\mathbf{Q}_j$ are the matrices whose entries are the $p_{j,ft}$ and $q_{j,ft}$ defined in~\eqref{eq:p_em} and~\eqref{eq:q_em}.
By applying exactly the same methodology, we obtain the update on $\textbf{H}$:
\begin{equation}
\mathbf{H}_j \leftarrow \mathbf{H}_j \odot \left( \frac{ \mathbf{W}_j^\mathsf{T} (\mathbf{P}_j \odot \mathbf{V}_j^{\odot -2})    }{\mathbf{W}_j^\mathsf{T} (\mathbf{V}_j^{\odot -1} + \mathbf{Q}_j \odot  \mathbf{V}_j^{\odot -3/2} )  }     \right)^{\odot 1/2}.
\label{eq:Hj_em}
\end{equation}

\subsubsection{Relation to other approaches}

We remark that if $\kappa = 0$, then $\lambda=\rho=0$: therefore, $q_{j,ft}=0$ and $p_{j,ft}$ becomes the posterior power of $s_{j,ft}$, as mentioned in Section~\ref{sec:estim_nmf_func}. Then, we recognize in~\eqref{eq:QMLnmf} the IS divergence between $\mathbf{P}_j$ and $\mathbf{W}_j \mathbf{H}_j$, as in the EM algorithm for ISNMF~\cite{Magron2018b}. Consequently, the updates rules~\eqref{eq:Wj_em} and~\eqref{eq:Hj_em} are similar to those obtained in such a scenario~\cite{Magron2018b}, up to an additional power $1/2$, which is common when applying the majorize-minimization methodology for estimating ISNMF~\cite{Fevotte2011a}.

Besides, one can consider an alternative AG model as described in Section~\ref{sec:model_relation}. If one considers that the sources are centered  ($s_j \sim \mathcal{N}(0,\Gamma_j)$), then $\mathbf{Q}_j = 0$: we recognize in~\eqref{eq:QMLnmf} the IS divergence between the NMF model and the phase-corrected posterior power. The derivation of the update rules is then easier than in the general case, since it eliminates the need for the majorize-minimization method: one can apply the commonly-used heuristic method described in~\cite{Lee1999} to obtain alternative multiplicative update rules. This approach is described in more details in~\cite{Magron2018e}.

\subsection{M-step: phase parameters}
\label{sec:estim_phase}

Let us now derive the updates on the phase parameters. We rewrite the functional~\eqref{eq:QMLlong} by removing the terms that do not depend on the phase parameters, which leads to:
\begin{equation}
\mathcal{Q}^{\text{ML}}(\Theta|\Theta^{(i-1)}) \overset{c}{=} \sum_{j=1}^{J} \sum_{f,t} \Re \left( \alpha_{j,ft} e^{-2 \mathrm{i} \mu_{j,ft} } +  \beta_{j,ft} e^{-\mathrm{i} \mu_{j,ft} } \right),
\end{equation}
with:
\begin{equation}
\alpha_{j,ft} = \frac{\rho}{( (1-\lambda^2)^2-\rho^2) [\mathbf{W}_j \mathbf{H}_j]_{ft}}  ( c'_{j,ft} + m'^2_{j,ft} ),
\end{equation}
and:
\begin{equation}
\beta_{j,ft} =  \frac{2  \lambda (1-\lambda^2-\rho) }{( (1-\lambda^2)^2-\rho^2) \sqrt{[\mathbf{W}_j \mathbf{H}_j]_{ft}} }  m'_{j,ft} .
\label{eq:beta_em}
\end{equation}
Therefore, adding the log-prior over the phase parameters~\eqref{eq:phaseprior} leads to maximizing the following functionals:
\begin{equation}
g_{j,ft}(\mu_{j,ft}) = \Re \left( \alpha_{j,ft} e^{-2 \mathrm{i} \mu_{j,ft} } + \tilde{\beta}_{j,ft} e^{- \mathrm{i} \mu_{j,ft} } \right),
\label{eq:g_phase}
\end{equation}
with respect to $\mu_{j,ft}$, and where:
\begin{equation}
\tilde{\beta}_{j,ft}= \beta_{j,ft} + \tau \left( e^{ \mathrm{i} \mu_{j,ft-1} + 2 \mathrm{i} \pi l \nu_{j,ft} } + e^{ \mathrm{i} \mu_{j,ft+1} - 2 \mathrm{i} \pi l \nu_{j,ft+1} } \right).
\label{eq:betatilde_em} 
\end{equation}
Let us remove the indexes $j,ft$ in what follows for more clarity. We then seek to maximize:
\begin{align}
g(\mu) &= \Re \left( \alpha e^{-2 \mathrm{i} \mu } + \tilde{\beta} e^{- \mathrm{i} \mu } \right) \\
   &= |\alpha| \cos (2\mu-\angle \alpha) + |\tilde{\beta}| \cos (\mu-\angle \tilde{\beta}), \label{eq:g_full}
\end{align}
which leads to finding the roots of:
\begin{equation}
g'(\mu) =- 2 |\alpha| \sin (2\mu-\angle \alpha) - |\tilde{\beta}| \sin (\mu-\angle \tilde{\beta}).
\end{equation}
Unfortunately, it is not straightforward to write the solutions of this problem in closed-form. Besides, it requires further operations to determine which root maximizes $g$, leading to a quite computationally intensive procedure. Instead, drawing on~\cite{Magron2018a}, since we experimentally observed that $|\alpha| << |\beta|$, we propose to approximate~\eqref{eq:g_full} by:
\begin{equation}
\tilde{g}(\mu) = \Re \left( \tilde{\beta} e^{-i \mu} \right)=|\tilde{\beta}| \cos (\mu -  \angle \tilde{\beta}),
\label{eq:g_approx}
\end{equation}
which is easily maximized by $\mu = \angle \tilde{\beta} $. This update depends on the values of the phase parameter in frames $t-1$ and $t+1$, so it has to be applied sequentially over time frames (which is common when using Markov chain priors such as in~\cite{Bertin2010}).

To assess the validity of this update scheme, we applied both procedures (maximization of the exact functional~\eqref{eq:g_full} and its approximation~\eqref{eq:g_approx}) on the learning dataset used in the experimental evaluation (see Section~\ref{sec:exp_protocol}). The average relative difference between the phases obtained with those two approaches was of approximately $10^{-5}$. Consequently, we propose to use the approximate update scheme, since it yields very similar estimates while being significantly faster than performing the exact maximization.

Finally, if one consider an alternative AG model with null relation terms (\textit{cf}. Section~\ref{sec:model_relation}), then $\alpha=0$, which eliminates the need for this simplifying assumption. It also modifies the values of $\beta$, $p$ and $q$, therefore leading to a different procedure, which will be investigated in future work.

\subsection{Full procedure}

The EM procedure is summarized in Algorithm~\ref{al:EM}.
The phase location parameters $\mu_j$ are initialized by assigning the mixture phase to each source. The initialization of the NMF matrices is discussed in Sections~\ref{sec:exp:proto:semi_iss} and~\ref{sec:exp_init}.

The frequencies $\nu$ are provided as inputs of the algorithm. We estimate them by means of a quadratic interpolated FFT (QIFFT)~\cite{Abe2004} on the log-spectra of the initial variance estimates $\textbf{V}_j$. This estimation is performed locally (at each time frame) in order to account for slow variations of the frequencies. The frequency range is then decomposed into \textit{regions of influence}~\cite{Magron2015a} to ensure that the phase in a given channel is unwrapped with the appropriate frequency. 

This algorithm includes a normalization step after updating $\mathbf{W}_j$ and $\mathbf{H}_j$, which eliminates trivial scale indeterminacies and avoids numerical instabilities. We impose a unitary $\ell_2$-norm on each column of $\mathbf{W}_j$ and scale $\mathbf{H}_j$ accordingly, so that the cost function is not affected.

Finally, one final E-step is performed after looping in order to estimate the sources with the most up-to-date parameters.

\begin{algorithm}[t]
	\caption{EM algorithm for complex ISNMF}
	\label{al:EM}
			\textbf{Inputs}: Mixture $\mathbf{X} \in \mathbb{C}^{F \times T}$,\\
            Phase parameters $\kappa$ and $\tau$ $\in \mathbb{R}_{+}$,\\
            Initial NMF matrices $\forall j$, $\mathbf{W}_j \in \mathbb{R}_{+}^{F \times K_j}$, $\mathbf{H}_j \in \mathbb{R}_{+}^{K_j \times T}$,\\
            Initial phases $\forall j$, $\mu_j \in [0,2\pi[^{F \times T}$,\\
            Normalized frequencies $\forall j$, $\nu_j \in \mathbb{R}^{\times F \times T}$.\\
			
			\textbf{Anisotropy parameters}:\\
            Compute $\lambda$ and $\rho$ with~\eqref{eq:ag_params}. \\

			\While{\upshape{stopping criterion not reached}}{
			
			\textcolor{gray}{\% E-step}
            
			Update $m$, $\gamma$ and $c$ with~\eqref{eq:ag_moments}, \\
			Update $m_x$, $\gamma_x$ and $c_x$ with~\eqref{eq:moments_X}, \\
			Update $m'$ with~\eqref{eq:posterior_mean}, \\
			Update $\gamma'$ and $c'$ with~\eqref{eq:posterior_var}, \\
			
			\textcolor{gray}{\% M-step: NMF}
            
			Update $p$ with~\eqref{eq:p_em} and $q$ with~\eqref{eq:q_em}. \\
			$\forall j$, Update $\mathbf{W}_j$ with~\eqref{eq:Wj_em} and $\mathbf{H}_j$ with~\eqref{eq:Hj_em}, \\
            Normalize $\mathbf{W}$ and $\mathbf{H}$.
            
            \textcolor{gray}{\% M-step: phase}
            
			Update $\beta$ with~\eqref{eq:beta_em}. \\
			
			\For{$t=1$ \KwTo $T-2$}{
			
			$\forall (j,f)$, update $\tilde{\beta}_{j,ft}$ with~\eqref{eq:betatilde_em}, \\
			$\mu_{j,ft} = \angle \tilde{\beta}_{j,ft}$. \\
			
			}

			}

            Update $m$, $\gamma$ and $c$ with~\eqref{eq:ag_moments}, \\
            Update $m_x$, $\gamma_x$ and $c_x$ with~\eqref{eq:moments_X}, \\
			Update $m'$ with~\eqref{eq:posterior_mean}. \\
			
			\textbf{Outputs}: $m' \in \mathbb{C}^{J \times F \times T}$.
		
\end{algorithm}

\section{Experimental evaluation}
\label{sec:exp}

In this section, we experimentally assess the potential of the proposed complex ISNMF model for a task of monaural musical source separation. Sound excerpts can be found on the companion website for this paper~\cite{cisnmf_webpage}. In the spirit of reproducible research, the code of this experimental study is available online\footnote{\url{https://github.com/magronp/complex-isnmf}}.

\subsection{Protocol}
\label{sec:exp_protocol}

\subsubsection{Dataset}

We consider $100$ music song excerpts from the DSD100 database, a semi-professionally mixed set of music songs used for the SiSEC 2016 campaign~\cite{Liutkus2017}. Each excerpt is $10$ seconds long and is made up of $J=4$ sources: \texttt{bass}, \texttt{drum}, \texttt{vocals} and \texttt{other}. The database is split into two subsets of $50$ songs: a learning set, on which the meta-parameters of the algorithms are tuned and the initialization strategies are investigated, and a test set, on which the separation benchmark is performed. The signals are sampled at $44100$ Hz and the STFT is computed with a $92$ ms long Hann window and $75$~\% overlap. The resulting STFTs are therefore matrices of dimensions $2049 \times 433$.

\subsubsection{Separation scenario}
\label{sec:exp:proto:semi_iss}

In \textit{coding-based informed source separation}~\cite{Liutkus2012a}, we assume some side-information can be computed from the isolated sources (the \textit{encoding} stage) and then used to perform separation (the \textit{decoding} stage). A common approach consists of computing a nonnegative matrix or tensor factorization~\cite{Rohlfing2016,Rohlfing2017,Ozerov2013} on the isolated source spectrograms and then using the corresponding decomposition to estimate a Wiener filter at the decoding stage. Here, we consider a \textit{semi-informed} scenario, in which the dictionaries $\textbf{W}_j$ are estimated on the isolated sources and the activation matrices $\textbf{H}_j$ computed from the mixture. This setting is less restrictive than a fully-informed setting since we only transmit the dictionaries instead of both NMF matrices. Note than since we use a learning dataset for tuning some parameters, this setting is actually supervised semi-informed, but we refer to it as semi-informed for brevity.

Dictionaries are learned with $200$ iterations of ISNMF applied to each isolated spectrogram, using multiplicative update rules~\cite{Fevotte2009}, random initial matrices and a rank of factorization $K_j=50$, which corresponds to an $8$-fold compression ratio. The dictionaries are then fixed at the separation stage, since we experimentally observed that it leads to better results than further updating them on the mixture.

\subsubsection{Comparison references}

As baselines, we test the consistent anisotropic Wiener (CAW) filter~\cite{Magron2017c} which combines the consistent~\cite{LeRoux2013} and anisotropic~\cite{Magron2017} Wiener filters, and we also consider the phase-constrained CNMF~\cite{Bronson2014,Magron2016,Rodriguez-Serrano2016}. In order to make the comparison fair, we implemented a version of CNMF known as CNMF with intra-source additivity~\cite{King2012}: it consists in modeling the phase $\phi_j$ of each source instead of the phase of each NMF component, as in the classical CNMF model~\cite{Kameoka2009}. This significantly reduces the number of parameters of the model, thus it lowers both the memory and computation time required for the estimation of the model, at the cost of a moderate drop in terms of separation quality~\cite{King2012}.

Source separation quality is measured with the signal-to-distortion, signal-to-interference, and signal-to-artifact ratios (SDR, SIR, and SAR)~\cite{Vincent2006} expressed in dB, where only a rescaling (not a refiltering) of the reference is allowed.

\begin{table*}[t]
	\center
    \caption{Source separation performance for each instrument (SDR, SIR and SAR in dB) averaged over the DSD100 test dataset.}
	\label{tab:results_separation_instruments}
	\begin{tabular}{l|ccc|ccc|ccc|ccc}
    \hline
    \hline
         & \multicolumn{3}{c|}{Bass} &  \multicolumn{3}{c|}{Drums} &  \multicolumn{3}{c|}{Other} &  \multicolumn{3}{c}{Vocals} \\
		 & SDR & SIR & SAR & SDR & SIR & SAR & SDR & SIR & SAR & SDR & SIR & SAR \\
		Wiener         & $2.6$  & $7.9$ & $4.4$ & $4.7$  & $17.4$ & $5.1$ & $3.7$  & $12.9$ & $4.4$ & $7.6$  & $18.1$ & $8.1$   \\
		AW             & $2.6$  & $8.1$ & $4.3$ & $4.4$  & $\textbf{18.5}$ & $4.7$ & $3.6$  & $\textbf{13.1}$ & $4.2$ & $7.5$  & $\textbf{18.9}$ & $7.9$ \\
        CAW            & $2.8$  & $8.1$ & $\textbf{4.5}$ & $4.8$  & $17.6$ & $5.1$ & $3.8$  & $12.9$ & $4.4$ & $7.0$  & $16.7$ & $7.5$  \\
        CNMF           & $2.3$  & $6.9$ & $4.5$ & $3.7$  & $12.8$ & $4.4$ & $2.6$  & $10.1$ & $3.7$ & $5.9$  & $15.7$ & $6.5$  \\
		Complex ISNMF  & $\textbf{3.0}$  & $\textbf{10.1}$ & $4.1$ & $\textbf{5.4}$  & $15.9$ & $\textbf{5.9}$ & $\textbf{3.8}$  & $12.4$ & $\textbf{4.6}$ & $\textbf{7.7}$  & $18.4$ & $\textbf{8.2}$  \\
       \hline
       \hline
	\end{tabular}
\end{table*}

\begin{table}[t]
	\center
    \caption{Source separation performance averaged over instruments: mean plus/minus standard deviation over the dataset.}
	\label{tab:results_separation_mean}
	\begin{tabular}{lccc}
    \hline
    \hline
		 & SDR & SIR & SAR \\
		Wiener         & $4.7 \pm 1.6$  		& $14.1 \pm 2.9$ 		  & $5.5 \pm 1.5$  \\
		AW             & $4.5 \pm 1.7$  		& $\textbf{14.6} \pm 2.8$ & $5.3 \pm 1.5$  \\
        CAW            & $4.6 \pm 2.0$  		& $13.8 \pm 2.7$ 		  & $5.4 \pm 2.0$  \\
        CNMF           & $3.6 \pm 1.7$  		& $11.4 \pm 2.3$ 		  & $4.8 \pm 1.6$  \\
		Complex ISNMF  & $\textbf{5.0} \pm 1.7$ & $14.2 \pm 2.8$ 		  & $\textbf{5.7} \pm 1.6$  \\
       \hline
       \hline
	\end{tabular}
\end{table}

\subsection{Initialization strategy}
\label{sec:exp_init}

We briefly investigate here on the best strategy for initializing the complex ISNMF algorithm at the separation stage, once the dictionaries are learned. A first approach is to provide a warm start to the algorithm thanks to $50$ iterations of ISNMF computed on the mixture, whose activation matrix is randomly initialized. Besides, it is necessary to have a first estimate of the variances in order to compute the frequencies, which are needed as inputs of Algorithm~\ref{al:EM}. On top of that initialization, we run $150$ iterations of complex ISNMF. Alternatively, we run $200$ iterations of complex ISNMF on top of a random initialization (though we still use the frequencies as computed before), so the total number of iterations is the same in both scenarios.

\begin{figure}[t]
	\centering
	\includegraphics[scale=0.5]{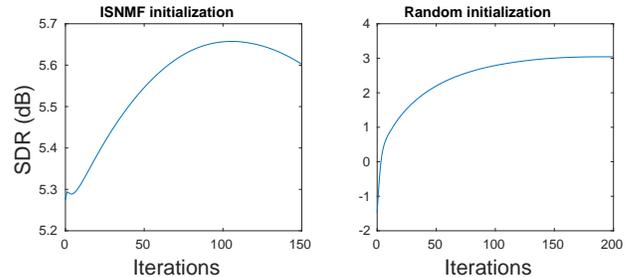}
	\caption{SDR over iterations for an ISNMF (left) and random (right) initialization.}
	\label{fig:init_strategy}
\end{figure}

We present the SDR over iterations in Fig.~\ref{fig:init_strategy} (results are averaged over the learning set) for $\kappa=\tau=0.5$: similar conclusions can be drawn from other values of the parameters and from the SIR and SAR. We observe that initializing complex ISNMF with ISNMF provides better results than a random initialization. Consequently, in the following experiments, we will retain this ISNMF-initialization strategy in order to bootstrap the complex ISNMF algorithm, which will use $100$ iterations.

\subsection{Phase parameters influence}
\label{sec:exp_ph}

We run the different methods on the $50$ songs that form the learning set in order to learn the optimal phase parameters.

\subsubsection{Complex ISNMF}

\begin{figure}[t]
	\hspace{-0.3cm}
	\includegraphics[scale=0.56]{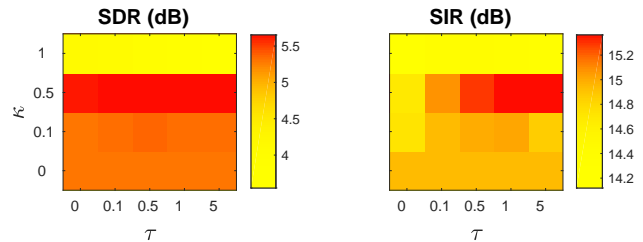}
	\caption{Influence of the phase parameters $\kappa$ and $\tau$ on the source separation quality (SDR and SAR are similar). The range is limited to $[0,1]$ and $[0,5]$ for $\kappa$ and $\tau$ respectively for clarity purpose, since the performance decreases outside of these ranges.}
	\label{fig:inf_phase}
\end{figure}

The results presented in Fig~\ref{fig:inf_phase} show that for non-null values of the phase parameters, the proposed approach can outperform a phase-unaware approach (for which $\kappa=\tau=0$) according to the SDR, SIR and SAR. We found that $\kappa = 0.5$ and $\tau = 5$ provides a quite good compromise between the different indicators.

\subsubsection{Phase-constrained CNMF}
This method depends on a weight parameter $\sigma_u$ which promotes the sinusoidal model phase constraint. The separation work flow is the same as for complex ISNMF, except we use here an NMF with Euclidean distance~\cite{Lee1999} for both dictionary learning and initialization on the mixture. Indeed, since CNMF is based on the Euclidean distance, learning IS-based dictionaries would not be consistent with the distortion metric in CNMF. The value $\sigma_u=10^{-2}$ appears as the best candidate, since the SDR is slightly reduced ($-0.2$ dB) compared to the unconstrained baseline (for which $\sigma_u=0$), but it allows for more interference reduction ($+1.4$ dB in SIR). Values of $\sigma_u$ greater than $10^{-2}$ still increase the SIR, but at the cost of a significant drop in SDR.

\subsubsection{Wiener filters}
CAW~\cite{Magron2017c} depends on two parameters $\kappa$ and $\delta$ which respectively promote anisotropy and consistency. We first estimate the variances with $150$ iterations of ISNMF on the mixture, and then we apply the filter. We propose the following sets of values:

\begin{itemize}
\item For $\kappa=1$ and $\delta=0$, the SIR is improved by $+0.6$ dB at the cost of a slight decrease in SDR ($-0.1$ dB) compared to the baseline Wiener filtering (for which $\kappa=\delta=0$). We simply refer to it as AW since the consistency weight is null in this setting.
\item For $\kappa = 0.1$ and $\delta = 10^{-3}$, the SIR is very slightly reduced compared to the baseline ($-0.02$ dB) while the SDR is increased by $0.05$ dB. We refer to it as CAW.
\end{itemize}

One may chose other values for the parameters in order to have the best possible SDR (or SIR/SAR), but the proposed settings yield an overall compromise which does not excessively favor one indicator over the others.

\subsection{Results of the benchmark}

We now consider the $50$ songs that form the test set and run the compared methods. The results for each instrumental source are presented in Table~\ref{tab:results_separation_instruments}, and the results averaged over instruments are presented in Table~\ref{tab:results_separation_mean}.

We observe that the proposed complex ISNMF approach yields the best results in terms of SDR and SAR for all instruments and among all the compared techniques, except for the \texttt{bass} track in terms of SAR. It also outperforms the phase-unaware Wiener filtering and the phase-constrained CNMF in terms of average SIR. This demonstrates the interest of exploiting some phase information in a probabilistic model to overcome the limitations of those baseline approaches, as stressed in the introduction of this paper.

The complex ISNMF estimates contain slightly more interference than the AW estimates (a $0.4$ dB difference in SIR on average), but less artifacts (a $0.4$ dB difference in SAR on average), which leads to a greater SDR. Therefore, it is overall preferable to employ this method than our preliminary approaches~\cite{Magron2017,Magron2017c} to perform a joint estimation of magnitude and phase.

Let us note that the metrics do not vary much from one technique to another. Indeed, the main difference between them is the phase recovery technique, which has less impact on the SDR, SIR and SAR than differences in terms of magnitude estimation strategy.

An informal perceptual evaluation is consistent with those results (sounds excerpts are available at~\cite{cisnmf_webpage}). In particular, CNMF introduces smearing artifacts in the separated sources, and the \texttt{bass} and \texttt{drum} tracks estimated with the Wiener filters are strongly corrupted by musical noise. In comparison, the proposed complex ISNMF method yields \texttt{bass} estimates which contain fewer artifacts and interference, and \texttt{drums} estimates with neater attacks.

\subsection{Fitting the data}

Finally, we investigate on the capability of the AG model to represent audio data, that is to say, to assess that the mixture variables $x_{ft}$ are well-represented by AG distributions. To do so, we need to normalize the variables $x_{ft}$ so that all TF entries become identically distributed, which allows us to compute their histogram, and therefore to compare their empirical and theoretical densities. Since $x_{ft} \sim \mathcal{N}(m_{x,ft},\Gamma_{x,ft})$, it can be shown that:
\begin{equation}
y_{ft} =   ( \underline{x}_{ft}-\underline{m}_{x,ft} )^\mathsf{H}   \Gamma_{x,ft}^{-1} ( \underline{x}_{ft}-\underline{m}_{x,ft} )
\label{eq:norm_AG}
\end{equation}
follows a chi-squared distribution with $2$ degrees of freedom~\cite{Picinbono1996}. Then, once the model is estimated, we compute the normalized variable $\textbf{Y}$ from the mixture $\textbf{X}$ according to~\eqref{eq:norm_AG}, and all the entries of $\textbf{Y}$ are expected to be identically chi-squared distributed. Finally, even if there are some dependencies between the $x_{ft}$ because of the NMF and phase models, they are conditionally independent given the model parameters, which are estimated beforehand in order to compute the $y_{ft}$ with~\eqref{eq:norm_AG}. The resulting variables $y_{ft}$ are then independent and identically distributed, thus it becomes possible to plot their histogram.

\begin{figure}[t]
	\centering
	\includegraphics[scale=0.6]{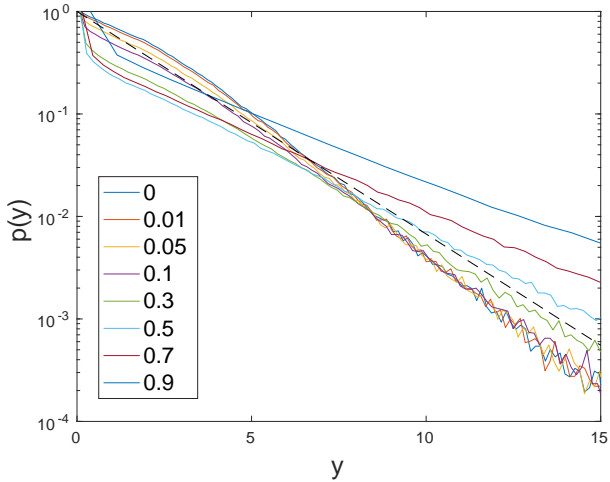}
	\caption{Empirical densities of the normalized data for several values of $\kappa$ (solid lines) and reference chi-squared density (dashed line).}
	\label{fig:chi2pdfs}
\end{figure}

The setting is the same as in the previous experiments, but we set $\tau$ at $0$ and we initialize Algorithm~\ref{al:EM} with the true phase values for $\mu_j$. Indeed, a fitting error can be due to a mismatch between the model and the observed data, but also to an estimation error. In this way, we only investigate on the accuracy of the model to represent the data, not on the phase estimation itself. The complex ISNMF algorithm is run on one song (similar results are obtained for the other songs) for several values of $\kappa$. The results are presented in Fig.~\ref{fig:chi2pdfs}.

We observe that small values of $\kappa$ lead to empirical densities that approach the theoretical one from above for small values of $x$ and from below for greater values of $x$. For greater values of $\kappa$, this trend is inverted. In particular, the value $\kappa=0.5$ leads to a good fit on average, which may explain why this value leads to the best results in terms of separation quality (see Section~\ref{sec:exp_ph}).

Overall, a better fit can be obtained with non-null values of $\kappa$, which demonstrates the interest of AG distributions over isotropic variables to represent audio data in the STFT domain.

\section{Conclusion}
\label{sec:conclu}

In this paper, we introduced complex ISNMF, a probabilistic model based on the AG distribution. It consists of modeling the sources with anisotropic random variables, which makes it possible to enforce some desirable phase properties, while classical circularly-symmetric variables do not allow one to favor a phase model. Therefore, it combines the advantages of ISNMF and CNMF, that is, using a distortion metric well adapted to audio and phase-awareness. We experimentally showed that it outperforms those two approaches, and thus appears as a good candidate for phase-aware audio source separation in semi-informed settings. This model is also suitable for supervised applications where some training material is available, but then it is required to account for the potential mismatch between training and test materials~\cite{Virtanen2009,Kitamura2013}.

An interesting direction for future work is the investigation of alternative phase-aware probabilistic models, in order to extend CNMF to other beta-divergences, as first attempted in~\cite{Kameoka2017}. Alternatively, one can exploit the family of multivariate stable distributions~\cite{Leglaive2017} with an  anisotropic shape matrix in order to combine phase-awareness and robust magnitude modeling~\cite{Magron2017b}. Finally, we could incorporate deep neural networks in this Bayesian framework for estimating the variances instead of using an NMF model, as it was done in a multichannel scenario with isotropic Gaussian variables~\cite{Nugraha2016}. Indeed, deep learning methods have shown remarkably good results for musical source separation~\cite{Takahashi2017}, but there is still some room for improvement, notably in terms of phase recovery, since those methods usually exploit a phase-unaware Wiener-like mask to estimate the complex-valued sources.

\appendix
\label{sec:anx_estep}

In this appendix, we detail the E-step of the proposed algorithm, which consists in computing the functional given by~\eqref{eq:QML}, which we recall hereafter:
\begin{equation*}
\mathcal{Q}^{\text{ML}}(\Theta,\Theta^{(i-1)}) = \int p(\textbf{S}|\textbf{X};\Theta^{(i-1)}) \log p(\textbf{X},\textbf{S};\Theta)  d\textbf{S}.
\end{equation*}
The complete data log-likelihood is given by:
\begin{align*}
&\log p(\textbf{X},\textbf{S};\Theta) = \sum_{f,t} \log p(x_{ft}|\textbf{s}_{ft};\Theta)  + \sum_{j=1}^{J'} \log p(s_{j,ft};\Theta) \\
&\overset{c}{=} -\frac{1}{2} \sum_{f,t} \log(|\Gamma_{J,ft}|) + B_{ft} + \sum_{j=1}^{J'} \log(|\Gamma_{j,ft}|) + A_{j,ft},
\end{align*}
where:
\begin{equation*}
A_{j,ft} = ( \underline{s}_{j,ft}-\underline{m}_{j,ft} )^\mathsf{H}   \Gamma_{j,ft}^{-1} ( \underline{s}_{j,ft}-\underline{m}_{j,ft} ),
\end{equation*}
and
\begin{equation*}
B_{ft} = ( \underline{x}_{ft} - \underline{m}_{J,ft}- \sum_{j=1}^{J'} \underline{s}_{j,ft} )^\mathsf{H}   \Gamma_{J,ft}^{-1} ( \underline{x}_{ft} - \underline{m}_{J,ft}- \sum_{j=1}^{J'} \underline{s}_{j,ft} ).
\end{equation*}
Therefore,~\eqref{eq:QML} rewrites:
\begin{multline}
\mathcal{Q}^{\text{ML}}(\Theta,\Theta^{(i-1)}) \overset{c}{=} -\frac{1}{2} \sum_{f,t}\sum_{j=1}^{J} \log(|\Gamma_{j,ft}|) \\
+   \sum_{f,t}\sum_{j=1}^{J'} \mathbb{E}_{\textbf{S}|\textbf{X};\Theta^{(i-1)}} \left( A_{j,ft}  \right) + \mathbb{E}_{\textbf{S}|\textbf{X};\Theta^{(i-1)}} \left( B_{ft} \right) .
\label{eq:QML_anx}
\end{multline}
Firstly, let us compute the expectation $\mathbb{E}_{\textbf{S}|\textbf{X};\Theta^{(i-1)}} \left( A_{j,ft} \right)$. We remove the indices $j,ft$ and the subscript ${\textbf{S}|\textbf{X};\Theta^{(i-1)}}$ for clarity. We have, thanks to the trace identity:
\begin{align*}
\mathbb{E}(A) &= \mathbb{E} \left(  ( \underline{s}-\underline{m} )^\mathsf{H}   \Gamma^{-1} ( \underline{s}-\underline{m} ) \right) \\
&= ( \underline{m}'-\underline{m} )^\mathsf{H}   \Gamma^{-1} ( \underline{m}'-\underline{m} ) + \text{Tr}(\Gamma^{-1} \Gamma' ).
\end{align*}
Besides, 
\begin{equation*}
\text{Tr}(\Gamma^{-1} \Gamma') = \frac{1}{|\Gamma|} ( \gamma \gamma' - \Re( \bar{c} c') ),
\end{equation*}
then:
\begin{equation*}
\mathbb{E} (A) = \frac{2}{|\Gamma|} \left( \gamma (|m'-m|^2 + \gamma') - \Re( \bar{c} ((m'-m)^2 + c') )  \right).
\end{equation*}
Now, let us compute $\mathbb{E}(B)$. We use, once again, the trace identity, which leads to:
\begin{align*}
&\mathbb{E}(B) = \mathbb{E} \left( ( \underline{x} - \underline{m}_{J}- \sum_{j=1}^{J'} \underline{s}_{j} )^\mathsf{H}   \Gamma_{J}^{-1} ( \underline{x} - \underline{m}_{J}- \sum_{j=1}^{J'} \underline{s}_{j} ) \right) \\
&= ( \underline{x} - \underline{m}_J- \sum_{j=1}^{J'} \underline{m}'_j )^\mathsf{H}   \Gamma^{-1} ( \underline{x} - \underline{m}_J - \sum_{j=1}^{J'} \underline{m}'_j ) + \text{Tr}(\Gamma_J^{-1} \Gamma'_J ).
\end{align*}
Thanks to the conservative property of the anisotropic Wiener filtering~\eqref{eq:posterior_mean}, we have $\sum_{j=1}^{J'} \underline{m}'_j = \underline{x} - \underline{m}'_J$, so:
\begin{equation*}
\mathbb{E}(B) = ( \underline{m}'_{J} - \underline{m}_{J} )^\mathsf{H}   \Gamma_{J}^{-1} ( \underline{m}'_{J} - \underline{m}_{J} ) + \text{Tr}(\Gamma_J^{-1} \Gamma'_J ). \\
\end{equation*}
Then, $\mathbb{E}(B)$ is similar to $\mathbb{E}(A)$, but applied to the last source $J$. Finally, incorporating the expressions of $\mathbb{E}(A)$ and $\mathbb{E}(B)$ into~\eqref{eq:QML_anx} leads to the expression of $\mathcal{Q}^{\text{ML}}$:
\begin{align*}
\mathcal{Q}^{\text{ML}}(\Theta,&\Theta^{(i-1)}) \overset{c}{=} -\sum_{f,t}\sum_{j=1}^{J} \log(\sqrt{|\Gamma_{j,ft}|})  \\
&+ \frac{1}{|\Gamma_{j,ft}|} \left( \gamma_{j,ft} (|m_{j,ft}'-m_{j,ft}|^2 + \gamma_{j,ft}')   \right) \\
&- \frac{1}{|\Gamma_{j,ft}|} \left( \Re( \bar{c}_{j,ft} ((m_{j,ft}'-m_{j,ft})^2 + c_{j,ft}') )  \right).
\end{align*}
%

\ifCLASSOPTIONcaptionsoff
  \newpage
\fi




\bibliographystyle{IEEEtran}
\bibliography{IEEEabrv,references}

\begin{thebibliography}{10}
\providecommand{\url}[1]{#1}
\csname url@samestyle\endcsname
\providecommand{\newblock}{\relax}
\providecommand{\bibinfo}[2]{#2}
\providecommand{\BIBentrySTDinterwordspacing}{\spaceskip=0pt\relax}
\providecommand{\BIBentryALTinterwordstretchfactor}{4}
\providecommand{\BIBentryALTinterwordspacing}{\spaceskip=\fontdimen2\font plus
\BIBentryALTinterwordstretchfactor\fontdimen3\font minus
  \fontdimen4\font\relax}
\providecommand{\BIBforeignlanguage}[2]{{%
\expandafter\ifx\csname l@#1\endcsname\relax
\typeout{** WARNING: IEEEtran.bst: No hyphenation pattern has been}%
\typeout{** loaded for the language `#1'. Using the pattern for}%
\typeout{** the default language instead.}%
\else
\language=\csname l@#1\endcsname
\fi
#2}}
\providecommand{\BIBdecl}{\relax}
\BIBdecl

\bibitem{Comon2010}
P.~Comon and C.~Jutten, \emph{Handbook of blind source separation: independent
  component analysis and applications}.\hskip 1em plus 0.5em minus 0.4em\relax
  Academic press, 2010.

\bibitem{Lee1999}
D.~D. Lee and H.~S. Seung, ``{Learning the parts of objects by non-negative
  matrix factorization},'' \emph{Nature}, vol. 401, no. 6755, pp. 788--791,
  1999.

\bibitem{Virtanen2007}
T.~Virtanen, ``Monaural sound source separation by nonnegative matrix
  factorization with temporal continuity and sparseness criteria,'' \emph{IEEE
  Transactions on Audio, Speech, and Language Processing}, vol.~15, no.~3, pp.
  1066--1074, March 2007.

\bibitem{Fevotte2009}
C.~F{\'e}votte, N.~Bertin, and J.-L. Durrieu, ``{Nonnegative matrix
  factorization with the {Itakura-Saito} divergence: With application to music
  analysis},'' \emph{Neural computation}, vol.~21, no.~3, pp. 793--830, March
  2009.

\bibitem{Virtanen2008}
T.~Virtanen, A.~T. Cemgil, and S.~Godsill, ``{Bayesian extensions to
  non-negative matrix factorisation for audio signal modelling},'' in
  \emph{{Proc. of IEEE International Conference on Acoustics, Speech and Signal
  Processing (ICASSP)}}, May 2008, pp. 1825--1828.

\bibitem{Liutkus2015a}
A.~Liutkus, D.~Fitzgerald, and R.~Badeau, ``Cauchy nonnegative matrix
  factorization,'' in \emph{{Proc. of IEEE Workshop on Applications of Signal
  Processing to Audio and Acoustics (WASPAA)}}, October 2015, pp. 1--5.

\bibitem{Simsekli2015}
U.~Simsekli, A.~Liutkus, and A.~T. Cemgil, ``Alpha-stable matrix
  factorization,'' \emph{IEEE Signal Processing Letters}, vol.~22, no.~12, pp.
  2289--2293, December 2015.

\bibitem{Fevotte2005}
C.~Fevotte and J.~F. Cardoso, ``Maximum likelihood approach for blind audio
  source separation using time-frequency {G}aussian source models,'' in
  \emph{Proc. of IEEE Workshop on Applications of Signal Processing to Audio
  and Acoustics (WASPAA)}, October 2005, pp. 78--81.

\bibitem{Liutkus2015}
A.~Liutkus and R.~Badeau, ``{Generalized Wiener filtering with fractional power
  spectrograms},'' in \emph{{Proc. of IEEE International Conference on
  Acoustics, Speech and Signal Processing (ICASSP)}}, April 2015, pp. 266--270.

\bibitem{Magron2015}
P.~Magron, R.~Badeau, and B.~David, ``{Phase recovery in NMF for audio source
  separation: an insightful benchmark},'' in \emph{Proc. of IEEE International
  Conference on Acoustics, Speech and Signal Processing (ICASSP)}, April 2015,
  pp. 81--85.

\bibitem{Parry2007}
R.~M. Parry and I.~Essa, ``Incorporating phase information for source
  separation via spectrogram factorization,'' in \emph{{Proc. of IEEE
  International Conference on Acoustics, Speech and Signal Processing
  (ICASSP)}}, April 2007, pp. II--661–II--664.

\bibitem{Kameoka2009}
H.~Kameoka, N.~Ono, K.~Kashino, and S.~Sagayama, ``{Complex {NMF}: A new sparse
  representation for acoustic signals},'' in \emph{{Proc. of IEEE International
  Conference on Acoustics, Speech and Signal Processing (ICASSP)}}, April 2009,
  p. 3437–3440.

\bibitem{LeRoux2009a}
J.~{Le Roux}, H.~Kameoka, E.~Vincent, N.~Ono, K.~Kashino, and S.~Sagayama,
  ``{Complex {NMF} under spectrogram consistency constraints},'' in
  \emph{{Proc. of Acoustical Society of Japan Autumn Meeting}}, September 2009.

\bibitem{Griffin1984}
D.~Griffin and J.~S. Lim, ``{Signal estimation from modified short-time
  {F}ourier transform},'' \emph{IEEE Transactions on Acoustics, Speech and
  Signal Processing}, vol.~32, no.~2, pp. 236--243, April 1984.

\bibitem{LeRoux2008}
J.~{Le Roux}, N.~Ono, and S.~Sagayama, ``{Explicit consistency constraints for
  {STFT} spectrograms and their application to phase reconstruction},'' in
  \emph{{Proc. of ISCA Workshop on Statistical and Perceptual Audition
  (SAPA)}}, September 2008, pp. 23--28.

\bibitem{McAuley1986}
R.~J. McAuley and T.~F. Quatieri, ``{Speech analysis/Synthesis based on a
  sinusoidal representation},'' \emph{IEEE Transactions on Acoustics, Speech
  and Signal Processing}, vol.~34, no.~4, pp. 744--754, August 1986.

\bibitem{Krawczyk2012}
M.~Krawczyk and T.~Gerkmann, ``{STFT} phase improvement for single channel
  speech enhancement,'' in \emph{Proc. of International Workshop on Acoustic
  Signal Enhancement (IWAENC)}, September 2012, pp. 1--4.

\bibitem{Magron2018}
P.~Magron, R.~Badeau, and B.~David, ``Model-based {STFT} phase recovery for
  audio source separation,'' \emph{{IEEE/ACM Transactions on Audio, Speech and
  Language Processing}}, vol.~26, no.~6, pp. 1095--1105, June 2018.

\bibitem{Krawczyk2014}
M.~Krawczyk and T.~Gerkmann, ``{STFT} phase reconstruction in voiced speech for
  an improved single-channel speech enhancement,'' \emph{IEEE/ACM Transactions
  on Audio, Speech, and Language Processing}, vol.~22, no.~12, pp. 1931--1940,
  December 2014.

\bibitem{Mowlaee2015}
P.~Mowlaee and J.~Kulmer, ``{Harmonic phase estimation in single-channel speech
  enhancement using phase decomposition and SNR information},'' \emph{IEEE/ACM
  Transactions on Audio, Speech, and Language Processing}, vol.~23, no.~9, pp.
  1521--1532, September 2015.

\bibitem{Magron2015a}
P.~Magron, R.~Badeau, and B.~David, ``{Phase reconstruction of spectrograms
  with linear unwrapping: application to audio signal restoration},'' in
  \emph{{Proc. of European Signal Processing Conference (EUSIPCO)}}, August
  2015, pp. 1--5.

\bibitem{Laroche1999}
J.~Laroche and M.~Dolson, ``{Improved phase vocoder time-scale modification of
  audio},'' \emph{IEEE Transactions on Speech and Audio Processing}, vol.~7,
  no.~3, pp. 323--332, May 1999.

\bibitem{Bronson2014}
J.~Bronson and P.~Depalle, ``{Phase constrained complex {NMF}: {Separating}
  overlapping partials in mixtures of harmonic musical sources},'' in
  \emph{{Proc. of IEEE International Conference on Acoustics, Speech and Signal
  Processing (ICASSP)}}, May 2014, pp. 7475--7479.

\bibitem{Rodriguez-Serrano2016}
F.~J. Rodriguez-Serrano, S.~Ewert, P.~Vera-Candeas, and M.~Sandler, ``A
  score-informed shift invariant extension of complex matrix factorisation for
  improving the separation of overlapped partials in music recordings,'' in
  \emph{{Proc. of IEEE International Conference on Acoustics, Speech and Signal
  Processing (ICASSP)}}, March 2016, pp. 61--65.

\bibitem{Magron2016}
P.~Magron, R.~Badeau, and B.~David, ``{Complex {NMF} under phase constraints
  based on signal modeling: application to audio source separation},'' in
  \emph{{Proc. of IEEE International Conference on Acoustics, Speech and Signal
  Processing (ICASSP)}}, March 2016, pp. 46--50.

\bibitem{Gray1980}
R.~Gray, A.~Buzo, A.~Gray, and Y.~Matsuyama, ``Distortion measures for speech
  processing,'' \emph{IEEE Transactions on Acoustics, Speech, and Signal
  Processing}, vol.~28, no.~4, pp. 367--376, August 1980.

\bibitem{Magron2017}
P.~Magron, R.~Badeau, and B.~David, ``{Phase-dependent anisotropic Gaussian
  model for audio source separation},'' in \emph{{Proc. of IEEE International
  Conference on Acoustics, Speech and Signal Processing (ICASSP)}}, March 2017,
  pp. 513--535.

\bibitem{Magron2018a}
P.~Magron and T.~Virtanen, ``{Bayesian anisotropic Gaussian model for audio
  source separation},'' in \emph{{Proc. of IEEE International Conference on
  Acoustics, Speech and Signal Processing (ICASSP)}}, April 2018, pp. 166 --
  170.

\bibitem{King2012b}
B.~King, C.~F{\'e}votte, and P.~Smaragdis, ``Optimal cost function and
  magnitude power for {NMF}-based speech separation and music interpolation,''
  in \emph{Proc. of IEEE International Workshop on Machine Learning for Signal
  Processing (MLSP)}, September 2012, pp. 1--6.

\bibitem{Agiomyrgiannakis2009}
Y.~Agiomyrgiannakis and Y.~Stylianou, ``{Wrapped Gaussian mixture models for
  modeling and high-rate quantization of phase data of speech},'' \emph{IEEE
  Transactions on Audio, Speech, and Language Processing}, vol.~17, no.~4, pp.
  775--786, May 2009.

\bibitem{Mardia1975}
K.~V. Mardia and P.~J. Zemroch, ``{Algorithm AS 86: The von Mises distribution
  function},'' \emph{Journal of the Royal Statistical Society. Series C
  (Applied Statistics)}, vol.~24, no.~2, pp. 268--272, 1975.

\bibitem{Gerkmann2014}
T.~Gerkmann, ``{MMSE-optimal enhancement of complex speech coefficients with
  uncertain prior knowledge of the clean speech phase},'' in \emph{{Proc. of
  IEEE International Conference on Acoustics, Speech and Signal Processing
  (ICASSP)}}, May 2014, pp. 4478--4482.

\bibitem{Gerkmann2014a}
------, ``Bayesian estimation of clean speech spectral coefficients given a
  priori knowledge of the phase,'' \emph{IEEE Transactions on Signal
  Processing}, vol.~62, no.~16, pp. 4199--4208, August 2014.

\bibitem{Watson1995}
G.~N. Watson, \emph{{A treatise on the theory of Bessel functions}}.\hskip 1em
  plus 0.5em minus 0.4em\relax Cambridge university press, 1995.

\bibitem{Picinbono1996}
B.~Picinbono, ``Second-order complex random vectors and normal distributions,''
  \emph{IEEE Transactions on Signal Processing}, vol.~44, no.~10, pp.
  2637--2640, October 1996.

\bibitem{Liutkus2018}
A.~Liutkus, C.~Rohlfing, and A.~Deleforge, ``{Audio source separation with
  magnitude priors: the BEADS model},'' in \emph{{Proc. of IEEE International
  Conference on Acoustics, Speech and Signal Processing (ICASSP)}}, April 2018,
  pp. 56 -- 60.

\bibitem{Beckmann1962}
P.~Beckmann, ``Statistical distribution of the amplitude and phase of a
  multiply scattered field,'' \emph{Journal of Research of the National Bureau
  of Standards}, vol. 66D, no.~3, pp. 231--240, May-June 1962.

\bibitem{LeRoux2013}
J.~{Le Roux} and E.~Vincent, ``Consistent {Wiener} filtering for audio source
  separation,'' \emph{IEEE Signal Processing Letters}, vol.~20, no.~3, pp.
  217--220, March 2013.

\bibitem{Bertin2010}
N.~Bertin, R.~Badeau, and E.~Vincent, ``Enforcing harmonicity and smoothness in
  {Bayesian} non-negative matrix factorization applied to polyphonic music
  transcription,'' \emph{IEEE Transactions on Audio, Speech and Language
  Processing}, vol.~18, no.~3, pp. 538--549, March 2010.

\bibitem{Dempster1977}
A.~P. Dempster, N.~M. Laird, and D.~B. Rubin, ``{Maximum likelihood from
  incomplete data via the EM algorithm},'' \emph{Journal of the royal
  statistical society. Series B (methodological)}, vol.~39, no.~1, pp. 1--38,
  1977.

\bibitem{Magron2017c}
P.~Magron, J.~Le~Roux, and T.~Virtanen, ``{Consistent anisotropic Wiener
  filtering for audio source separation},'' in \emph{{Proc. of IEEE Workshop on
  Applications of Signal Processing to Audio and Acoustics (WASPAA)}}, October
  2017, pp. 269--273.

\bibitem{Hunter2004}
D.~R. Hunter and K.~Lange, ``{A tutorial on {MM} algorithms},'' \emph{The
  American Statistician}, vol.~58, no.~1, pp. 30--37, 2004.

\bibitem{Fevotte2011}
C.~F{\'e}votte and J.~Idier, ``{Algorithms for nonnegative matrix factorization
  with the beta-divergence},'' \emph{Neural Computation}, vol.~23, no.~9, pp.
  2421--2456, September 2011.

\bibitem{Fevotte2011a}
C.~F{\'e}votte, ``{Majorization-minimization algorithm for smooth Itakura-Saito
  nonnegative matrix factorization},'' in \emph{Proc. of IEEE International
  Conference on Acoustics, Speech and Signal Processing (ICASSP)}, May 2011,
  pp. 1980--1983.

\bibitem{Lefevre2011}
A.~Lef{\`e}vre, F.~Bach, and C.~F{\'e}votte, ``{Itakura-Saito nonnegative
  matrix factorization with group sparsity},'' in \emph{Proc. of IEEE
  International Conference on Acoustics, Speech and Signal Processing
  (ICASSP)}, May 2011, pp. 21--24.

\bibitem{Magron2018b}
P.~Magron and T.~Virtanen, ``Expectation-maximization algorithms for
  {Itakura-Saito} nonnegative matrix factorization,'' in \emph{{Proc. of
  Interspeech}}, September 2018, pp. 856--860.

\bibitem{Magron2018e}
------, ``{Towards complex nonnegative matrix factorization with the
  beta-divergence},'' in \emph{{Proc. of the International Workshop on Acoustic
  Signal Enhancement (iWAENC)}}, September 2018.

\bibitem{Abe2004}
M.~Abe and J.~O. Smith, ``{Design criteria for simple sinusoidal parameter
  estimation based on quadratic interpolation of {FFT} magnitude peaks},'' in
  \emph{{Audio Engineering Society Convention 117}}, May 2004.

\bibitem{cisnmf_webpage}
\url{http://www.cs.tut.fi/~magron/demos/demo_CISNMF.html}.

\bibitem{Liutkus2017}
A.~Liutkus, F.-R. St{\"o}ter, Z.~Rafii, D.~Kitamura, B.~Rivet, N.~Ito, N.~Ono,
  and J.~Fontecave, ``The 2016 signal separation evaluation campaign,'' in
  \emph{{Proc. of International Conference on Latent Variable Analysis and
  Signal Separation (LVA/ICA)}}, February 2017, pp. 323--332.

\bibitem{Liutkus2012a}
A.~Liutkus, J.~Pinel, R.~Badeau, L.~Girin, and G.~Richard, ``{Informed source
  separation through spectrogram coding and data embedding},'' \emph{{Signal
  Processing}}, vol.~92, no.~8, pp. 1937--1949, 2012.

\bibitem{Rohlfing2016}
C.~Rohlfing, J.~M. Becker, and M.~Wien, ``{NMF}-based informed source
  separation,'' in \emph{Prof. IEEE International Conference on Acoustics,
  Speech and Signal Processing (ICASSP)}, March 2016, pp. 474--478.

\bibitem{Rohlfing2017}
C.~Rohlfing, J.~E. Cohen, and A.~Liutkus, ``Very low bitrate spatial audio
  coding with dimensionality reduction,'' in \emph{Prof. IEEE International
  Conference on Acoustics, Speech and Signal Processing (ICASSP)}, March 2017,
  pp. 741--745.

\bibitem{Ozerov2013}
A.~Ozerov, A.~Liutkus, R.~Badeau, and G.~Richard, ``Coding-based informed
  source separation: Nonnegative tensor factorization approach,'' \emph{IEEE
  Transactions on Audio, Speech, and Language Processing}, vol.~21, no.~8, pp.
  1699--1712, Aug 2013.

\bibitem{King2012}
B.~J. King, ``New methods of complex matrix factorization for single-channel
  source separation and analysis,'' Ph.D. dissertation, University of
  Washington, 2012.

\bibitem{Vincent2006}
E.~Vincent, R.~Gribonval, and C.~F{\'e}votte, ``Performance measurement in
  blind audio source separation,'' \emph{IEEE Transactions on Speech and Audio
  Processing}, vol.~14, no.~4, pp. 1462--1469, July 2006.

\bibitem{Virtanen2009}
T.~Virtanen and A.~T. Cemgil, ``Mixtures of gamma priors for non-negative
  matrix factorization based speech separation,'' in \emph{{Proc. of
  International Conference on Latent Variable Analysis and Signal Separation
  (LVA/ICA)}}, March 2009, pp. 646--653.

\bibitem{Kitamura2013}
D.~Kitamura, H.~Saruwatari, K.~Shikano, K.~Kondo, and Y.~Takahashi, ``Music
  signal separation by supervised nonnegative matrix factorization with basis
  deformation,'' in \emph{Proc. of International Conference on Digital Signal
  Processing (DSP)}, July 2013, pp. 1--6.

\bibitem{Kameoka2017}
H.~Kameoka, H.~Kagami, and M.~Yukawa, ``{Complex NMF with the generalized
  Kullback-Leibler divergence},'' in \emph{Proc. of IEEE International
  Conference on Acoustics, Speech and Signal Processing (ICASSP)}, March 2017,
  pp. 56--60.

\bibitem{Leglaive2017}
S.~Leglaive, U.~Simsekli, A.~Liutkus, R.~Badeau, and G.~Richard, ``Alpha-stable
  multichannel audio source separation,'' in \emph{Proc. of IEEE International
  Conference on Acoustics, Speech and Signal Processing (ICASSP)}, March 2017,
  pp. 576--580.

\bibitem{Magron2017b}
P.~Magron, R.~Badeau, and A.~Liutkus, ``{L\'evy NMF for robust nonnegative
  source separation},'' in \emph{{Proc. of IEEE Workshop on Applications of
  Signal Processing to Audio and Acoustics (WASPAA)}}, October 2017, pp.
  259--263.

\bibitem{Nugraha2016}
A.~A. Nugraha, A.~Liutkus, and E.~Vincent, ``Multichannel audio source
  separation with deep neural networks,'' \emph{IEEE/ACM Transactions on Audio,
  Speech, and Language Processing}, vol.~24, no.~9, pp. 1652--1664, September
  2016.

\bibitem{Takahashi2017}
N.~Takahashi and Y.~Mitsufuji, ``Multi-scale multi-band {DenseNets} for audio
  source separation,'' in \emph{{Proc. of IEEE Workshop on Applications of
  Signal Processing to Audio and Acoustics (WASPAA)}}, October 2017, pp.
  21--25.

\end{thebibliography}

\end{document}